\documentclass[conference]{IEEEtran}
\usepackage[utf8]{inputenc}
\usepackage{amsmath}
\usepackage{amssymb}
\usepackage{amsfonts}
\usepackage{algorithmic}
\usepackage{graphicx}
\usepackage{textcomp}
\usepackage{xcolor}
\usepackage[absolute]{textpos}

\usepackage{hyperref}
\hypersetup{pdfborder={0 0 0},colorlinks=true,urlcolor=blue,linkcolor=blue,citecolor=blue,bookmarks=true}
\hypersetup{pdftitle={Towards Implementing Responsible AI}}
\hypersetup{pdfauthor={Conrad Sanderson, Qinghua Lu, David Douglas, Xiwei Xu, Liming Zhu, Jon Whittle}}

\usepackage[shortlabels]{enumitem}
\usepackage{float}
\usepackage{color}
\usepackage{booktabs}
\usepackage{balance}

\makeatletter
\def\footnoterule{\kern-3\p@
  \hrule \@width 2in \kern 2.6\p@} 
\makeatother

\begin{document}

\title{Towards Implementing Responsible AI}

\author
  {
  Conrad Sanderson, Qinghua Lu, David Douglas, Xiwei Xu, Liming Zhu, Jon Whittle\\
  ~\\
  \textit{Data61 / CSIRO, Australia}
  }

\maketitle

\begin{abstract}
As the deployment of artificial intelligence (AI) is changing many fields and industries,
there are concerns about AI systems making decisions and recommendations without adequately considering various ethical aspects,
such as accountability, reliability, transparency, explainability, contestability, privacy, and fairness.
While many sets of AI ethics principles have been recently proposed that acknowledge these concerns,
such principles are high-level
and do not provide tangible advice on how to develop ethical and responsible AI systems. 
To gain insight on the possible implementation of the principles,
we conducted an empirical investigation involving semi-structured interviews with a cohort of AI practitioners.
The salient findings cover four aspects of AI system design and development,
adapting processes used in software engineering:
(i)~high-level view,
(ii)~requirements engineering,
(iii)~design and implementation,
(iv)~deployment and operation.
\end{abstract}

~


\begin{textblock}{13.4}(1.3,14.9)
\hrule
\vspace{1ex}
\noindent
\footnotesize
\textbf{{$^\ast$}~Published in:} IEEE International Conference on Big Data, pp. 5076--5081, 2022. DOI:~\href{https://doi.org/10.1109/BigData55660.2022.10021121}{10.1109/BigData55660.2022.10021121}
\end{textblock}

~

\section{Introduction}
\label{sec:intro}
\vspace{1ex}

Artificial intelligence (AI), which includes machine learning (ML),
can be helpful in solving many real-world problems,
and is hence changing many industries, especially in data rich domains~\cite{Hajkowicz_2023}.
In turn, the wide deployment of AI can greatly impact society at large.
However, there are many concerns about following decisions and recommendations made by AI systems,
which may not consider ethical aspects
such as accountability, reliability, transparency, explainability, contestability, privacy and fairness~\cite{Christoforaki_2022,Eubanks_2019,zhang21}.

Many high-level AI ethics principles and guidelines for responsible AI have been recently
proposed as an initial step to address the concerns~\cite{fjeld2020principled,jobin2019global}.
Across the various sets of principles proposed throughout the world,
a degree of consensus has been achieved~\cite{fjeld2020principled}.
Specifically, an analysis of 36 sets of AI ethics principles identified eight key themes:
privacy, accountability, safety and security, transparency and explainability,
fairness and non-discrimination, human control of technology,
promotion of human values,
and professional responsibility (which also covers responsible design, multi-stakeholder collaboration, and long-term impacts)~\cite{fjeld2020principled}.

An important limitation of such AI ethics principles is the lack of empirically proven methods
to robustly translate the principles into practice~\cite{Mittelstadt_2019,Sanderson_2023,Smit_2020}.
This includes the lack of methods that support the proactive design of transparent, explainable and accountable systems,
in contrast to the less useful post-hoc explanations of system outputs.
Existing methods overemphasise AI `explicability'
(which is necessary but not sufficient for transparency and \mbox{explainability}),
and are skewed to assessing the impact of an AI system on individuals rather than on society or groups.
Finally, the examined methods were found to be difficult to implement and typically positioned as discourse aids to document design decisions~\cite{Morley_2020}.

AI ethics can be considered as belonging to the broader field of computer ethics,
which seeks to describe and understand moral behaviour in creating and using computing technology~\cite{Whittle19}.
There is considerable overlap between the two:
non-AI software may also contain biases, infringe individual privacy, and be used for harmful purposes~\cite{friedman_bias_1996,johnson_ethics_2009}.
However, AI ethics can be distinguished from the regular ethical issues associated with software development 
by the decision making capabilities of AI systems (which can be opaque),
and the ability of some AI systems to continuously learn from input data.

The implementation of ethics principles in practice
requires an improved understanding of the practices of designers and developers of AI systems,
and how they relate to high-level ethics principles.
To this end, 
we have conducted an empirical investigation consisting of semi-structured interviews
with a cohort of scientists and engineers that develop and/or use AI/ML technologies across a wide range of projects.
The cohort of interviewees is from Australia's national science agency,
the Commonwealth Scientific and Industrial Research Organisation (CSIRO).
The agency is tasked with delivering scientific research across a diverse portfolio,
which includes a growing focus on the implementation of AI into many research and application areas
(such as health, agriculture, environment, biosecurity, etc).

We asked the interviewees what ethical issues they have considered in their AI/ML work,
and how they addressed or envisioned to address these issues.
The AI ethics principles proposed by the Australian government~\cite{DISER_2020} were treated
as a representative of the many similar principles from around the world~\cite{fjeld2020principled,jobin2019global},
and were used as a framing structure for the interviews, analysis and discussion.
The analysis of the interviews focuses on AI system design and development.
The observations, insights and challenges obtained from the interviews 
are discussed and placed in a wider context, 
with the aim of providing suggestions and caveats for
consideration when implementing high-level AI ethics principles.

The paper is continued as follows.
Section~\ref{sec:methods} details the methods used in the investigation.
Section~\ref{sec:analysis} analyses and discusses the interviewees' responses.
Limitations of this work are briefly discussed in Section~\ref{sec:limitations}.
Concluding statements are given in Section \ref{sec:conclusion}.

\section{Methods}
\label{sec:methods}

Semi-structured interviews were conducted with a cohort of 21 CSIRO scientists and engineers 
that use AI technologies (such as machine learning) within their projects.
The `use' of AI technologies in this context is defined as
the design, development, and/or implementation of systems with AI components.

The participants were initially recruited through a `call for participation' distributed across the organisation. 
This was followed by a snowballing technique where additional participants
were  sought via recommendations made by preceding participants,
until a saturation of perspectives from across the organisation was reached~\cite{Colvin_2016,Hall_2015}.
The main selection criterion for participants was self-reported involvement with research 
and/or development work that substantially included the use of AI methods and/or technologies.
We note that the selection and availability of interviewees may 
have constrained the nature and span of perspectives and outcomes of the study;
see Section~\ref{sec:limitations} for more details.

The interviewees had various backgrounds (computer science, engineering, health, biosecurity, natural resource management),
with a large variation in the interviewees' degree of experience and responsibility. 
10~interviewees worked primarily in computer science,
6~worked in the health \& biosecurity area,
and 5~worked in the land \& water area.
The gender split was approximately 76\% male and 24\% female.

Prior to each interview, each participant was provided with a summarised version
of the voluntary high-level ethics principles for AI
proposed by the Australian Government~\cite{DISER_2020};
see Fig.~\ref{fig:principles_summary} for details.
The interview protocol for each interview aimed to initially elicit
a subset of the high-level principles that was most relevant to each participant,
as experienced through their work.
The top 3-4 principles, as selected by the participant, were then explored via questions such as:
{(1)}~how each selected principle manifested itself in their work,
{(2)}~how each selected principle was addressed (using tools and/or processes),
{(3)}~what tools/processes would be useful in addressing each selected principle.
Follow-up questions aimed to cover relevant intersections with the following areas:
machine learning, software development, and ethics in AI.
Finally, participants were asked to reflect on other ethical considerations
or dilemmas not covered by the high-level principles but encountered and possibly addressed in their work.
All questions were posed in a conversational setting by three interviewers
that had diverse research backgrounds.
The interviews ranged from approximately 22 to 59 minutes in length, 
with a median length of approximately 37 minutes.

The investigation was approved by an internal ethics review committee,
in accordance with the National Statement on Ethical Conduct in Human Research~\cite{NMHRC_2018}.
The approval covered the scope of the investigation, the selection and contacting of potential participants,
the interview protocol, and the requirements for de-identification in transcripts.
Informed consent was sought and gained from all participants before their interviews.

~

The interview transcripts were independently analysed by the three interviewers,
using either thematic analysis~\cite{Braun_2006} or open card sorting~\cite{Paul_2008} 
to identify categories in the transcripts.
Codebook methods were used for the thematic analysis, 
where themes are understood as `domain summaries' and described before commencing analysis,
and the codes developed during the analysis of the transcripts are assigned to these themes~\cite{Braun_2006}. 
The eight high-level principles were used as domain summaries. 
The analysis was performed at a semantic level,
meaning that the analysis focused on describing and interpreting patterns identified in the interview data
rather than searching for any underlying assumptions or concepts within it~\cite{Braun_2006}.
The analyses derived from the interview transcripts were then cross-checked independently
by the three interviewers (to ensure inter-rater reliability and consensus on themes),
and finally combined into one overall analysis.


\begin{figure}[!b]
\centering
\fbox{%
\centering
\begin{minipage}{0.96\columnwidth}
\normalsize
\begin{enumerate}[(1),leftmargin=*]
\itemsep=1ex
\vspace{1ex}
\item
{\bf Privacy Protection \& Security.}
AI systems should respect and uphold privacy rights and regulations, and ensure the security of data.

\item
{\bf Reliability \& Safety.}
AI systems should operate reliably in accordance with their intended purpose during their lifecycle.

\item
{\bf Transparency \& Explainability.}
\textit{Transparency:} 
to ensure people know when they are being impacted by an AI system,
there should be transparency and responsible disclosure;
furthermore, people should be able to ascertain when they are engaging with an AI system;
\textit{Explainability:}
what the AI system is doing and why, including the employed data and processes.

\item
{\bf Fairness.}
AI systems should be inclusive and accessible, and should not involve or result in unfair discrimination
against individuals, communities or groups.

\item
{\bf Contestability.}
When an AI system significantly impacts a person, community, group or environment,
there should be a timely process that allows challenging the use and/or output of the system.

\item
{\bf Accountability.}
The people and/or organisations responsible for the various phases of the AI system lifecycle
should be identifiable and accountable for the outcomes of the system;
human oversight of AI systems should be enabled.

\item
{\bf Human, Social \& Environmental (HSE) Wellbeing.}
AI systems should benefit individuals, society, and the environment.

\item
{\bf Human-centered Values.}
AI systems should respect human rights, diversity, and the autonomy of individuals.
\vspace{1ex}
\end{enumerate}
\end{minipage}
} 
\caption
  {
  Summary of the voluntary high-level AI ethics principles proposed by the Australian Government~\cite{DISER_2020}.
  }
\label{fig:principles_summary}
\end{figure}

\newpage
\renewcommand{\baselinestretch}{0.99}\small\normalsize

\section{Analysis and Discussion}
\label{sec:analysis}

Results are reported for categories that were identified as being relevant to the development process of AI systems:
{(A)}~high-level view,
{(B)}~requirements engineering,
{(C)}~design and implementation,
{(D)}~deployment and operation.
The discussion on each part is further divided into salient points.
For each category, observations and insights are stated, and where appropriate, relevant comments from the interviewees are presented.
Quotations and paraphrased sentences are attributed to specific interview participants via markers in the form of (X\#\#),
where {\#\#} denotes a two-digit interviewee identifier.
In the text below, by `model' we mean a machine learning model (eg.~a neural network architecture) and its parameters.

\subsection{High-Level View}

\textbf{Assessment of ethical risk.} 
Throughout the interviews, many types of ethical risks were discussed
and approached in various ways.
As an example, for ensuring fairness the incomplete data problem was noted:
{\it ``you can be limited in what data [you have] available to use in the first place''}~(X01).
However, ethical risks were typically considered and checked in isolation, and were mostly around data and ML models.
We further observed that while rudimentary ethical risk assessment frameworks were used in practice,
they were not specifically tailored to AI system development:
\textit{``there was a privacy impact assessment; we went through a lengthy process to understand the privacy concerns 
and built in provisions to enable privacy controls''} (X10).
These types of risk assessments generally follow a do-once-and-forget approach,
which does not take into account AI systems that may continually learn and adapt.
It was also argued that adherence to the \textit{Transparency \& Explainability} principle can be merely an interim target related to operational risk:
\textit{``once I know that [the system] works most of the time I don't need explainability [and] transparency.
It's just temporary to establish the risk profile''} (X11).
Overall, it appears there is a lack of a comprehensive system-level checklist that covers all relevant ethics aspects throughout the full lifecycle of AI systems.

\textbf{Trust.}
Many interviewees acknowledged the importance of human trust in AI;
for example:
\textit{``a~lot of the work that we do trust comes as an important factor here,
that a user [...] who takes that information, wants to be able to trust it''} (X09).
Gaining and then maintaining trust from the providers of data that is used to train the AI system 
was identified as an important factor (and potentially an obstacle) for the development of reliable AI systems~(X02).
Contestability can contribute to trust:
\textit{``it can be very hard to get people to trust an analytical system that is just telling them to do something
and does not give them the choice to disagree with the system''} (X15). 
Furthermore, evidence needs to be presented to enable trust by humans (X12,~X18).
The evidence can be in forms such as demonstrated reliable operation, and explainability of the results produced by an AI system (X21).

\textbf{Credentials.} 
Several interviewees suggested that by attaching ethics credentials to AI components and products can enable a degree of responsible AI;
for example: \textit{``Getting those certificates, it always helps. As long as there is standardisation around it.''} (X13).
Even partial certification can be useful, related to the underlying hardware used by AI systems:
\textit{``A~lot of hardware is actually certified. [In] full size aviation, you have at least a certification.
So when you buy something you get some sort of guarantees''} (X12).

\textbf{Development type.} 
Two development types were often mentioned in the interviews: 
requirements-driven and outcome-driven, as well as a mixture of the two~\cite{jan2019}.
An iterative approach to outcome-driven development was mentioned:
\textit{``[development] is a continual and [iterative] process: humans need to continually evaluate the performance, identify [gaps] and provide insight into what's missing.
Then go back to connect data and refine the model''} (X02).

\textbf{System-level development tools.}
The significance of a system-level approach to AI development was recognised by several interviewees;
for example:
\textit{``[the] AI was designed and deployed as an end-to-end solution,
it wasn't that AI sat in the middle [...] it actually had to sit within the system''} (X14).
Lack of tools to help with addressing ethics principles was mentioned.
For example, manual work is currently required avoid accidental collection and use of sensitive data:
\textit{``we had to go through a lot of data and make sure that there was not a single frame with a person in it''} (X13).

\subsection{Requirements Engineering}

\textbf{Ethics requirements.}
Privacy and security were the most discussed requirements across the interviews.
Some principles (eg.~\textit{HSE Wellbeing}),
were often expressed only as indirect objectives instead of verifiable requirements and outcomes:
\textit{``the project leader might frame the project with we're working on improving [grass species] yield forecasting using machine learning.
You do feel good about working on projects that provide environmental benefit''} (X09).

Rather than relying purely on software engineers,
ethics requirements may need to be analysed and verified by a range of specialists and domain experts.
It was noted that AI system developers may opt to seek legal advice
to confirm that an AI system is following existing legal rules in a given application domain (X06).
In some cases, clients were unaware of privacy requirements regarding use of personally identifiable information:
\textit{``we had to contact our privacy officer to [...] confirm that's the case,
then we had to escalate that to the client, to let them know of that potential issue''} (X10).

To address the \textit{Reliability \& Safety} principle,
the approach of fail-safe by design was recommended,
although with the caveat that
\textit{``there's only so much you can think ahead about what those failure modes might be''} (X16).

\textbf{Responsibility scope.}
There were various meanings and interpretations of responsible AI.
The interviewees considered the following three interpretations~\cite{Tigard2021} as important:
normative (ie.~behaving in positive and socially acceptable forms),
possessive (ie.~duty and obligation),
descriptive (ie.~worthy of response/answerable).
The exact meaning of responsibility in the context of autonomous aerial systems was unclear:
\textit{``what happens if [the] remote pilot is really there, flicks the [disable] switch and the system doesn't react?
The remote pilot is not always in full control of [the UAV] because of technical reasons [like a failed communications link]''} (X12).
Moreover, the temporal span of responsibility may also need to be taken into account:
\textit{``whether the stuff works in 10 years, it's not under our control''} (X11).

\subsection{Design and Implementation}

\textbf{Incorporation of AI.}
Incorporating AI into a system can be a major architectural decision during the system design process.
A closely related design decision is whether the system allows users to make the final judgements,
rather than purely relying on the AI component.
This may involve allowing the AI component to be optionally disabled,
or changed from decision mode to suggestion mode.
For example, overriding AI provided decisions in medical contexts is seen as important:
\textit{``there was actually a defined process where if a patient was not flagged as being high risk,
[...] clinicians were still allowed to include the patient into the next step clinical review''} (X18).
Arguments against incorporating AI into systems include reduced interpretability
due to the complexity and/or vastness of machine learning models:
\textit{``traditionally, in statistics, people have used simpler linear models and that kind of thing.
They really worry about the parameters and they assign meaning to the parameters''} (X21).

\textbf{Trade-offs.}
Many tensions and trade-offs were noted between various ethics principles.
For example, reliability is in tension with fairness:
\textit{``we are in the spot where by design we restrict the variance as much as possible to make it easier to find a signal''} (X11).
Reliability is also in tension with privacy:
\textit{``if you [have] other ways of protecting privacy that don't involve aggregating, then you can be actually getting better distributional properties''} (X01).
The reliability of AI can greatly depend on the quantity and quality of the training data:
\textit{``if you're training a model without a lot of data, you can actually get some really weird results''} (X09).
Obtaining a sufficient number of samples to ensure reliability can be challenging,
as in some contexts (such as genomics) acquiring even one sample can be high in terms of financial and/or time costs, and may also involve privacy issues (X03).

In many trade-off cases, one principle was chosen in favour of other principles,
in contrast to building balanced trade-offs between principles,
where a cohort of stakeholders collectively evaluates value and risk~\cite{Whittlestone_2019}.
In cases involving tensions between privacy and reliability, federated learning was suggested:
\textit{``[various] research institutions from around the world can collaborate, because they don't have to give up their data;
they don't have to share their data''} (X03).

\textbf{Reuse.}
The reuse of trained AI models and related components was desired,
since training models and building components from scratch can be costly and/or time-consuming.
Furthermore, there was also a desire to reuse and/or iteratively adapt the overall design and architecture of existing AI systems, in order to allow training with new datasets~(X13).
The downside of the such reuse and adaptation includes accumulation of technical debt over time,
leading to increased maintenance issues~\cite{google_technical_debt},
which in turn may affect reliability.

\textbf{Explainability.}
Interviewees considered practical aspects of explainability and interpretability
by adopting human-centered approaches that take into account the background, culture, and preferences of users~\cite{Arrieta_2020}.
Users are more likely to trust the recommendations made by AI systems if there is supporting evidence for a given prediction/recommendation:
\textit{``there have been instances where we've chosen an explainable model which has slightly [lower] performance
[than] a non-explainable model which has higher performance but would be harder to convey the reasoning behind the prediction''} (X18).
Explainability was also seen as a waypoint to establish trust:
\textit{``[explainability] is just a temporary thing until people know it works''} (X11).
Explainability was often discussed in terms of the interface design:
\textit{``[...] nobody seems to ask about, what's the predictive performance of the algorithm [in the initial stakeholder meeting]?
[Instead] can I look at your interface and [...] see a couple of patient risk profiles and then understand that''} (X18).

\subsection{Deployment and Operation}

\textbf{Continuous validation.}
To ensure adherence to ethics requirements, continuous monitoring and validation of AI systems post-deployment was seen as necessary:
\textit{``it's up to us to come with technology that makes it acceptable for them to implement measurements [...] and being able to prove compliance''} (X07). 
Furthermore, awareness of potential mismatches between training data and data seen in operation is necessary to ensure AI models are used for their intended purpose (X04).
For maintaining the reliability of AI systems over the long term, model updates and retraining on newer and/or more comprehensive data were noted as important:
\textit{``If you build a model on 10 year old data, then you're not representing the current state of risks for certain disease.
As a minimum, [recalibration] on new data would probably be more meaningful''} (X18).

\textbf{Traceability of artefacts.}
Two main approaches were often identified related to traceability, provenance and reproducibility:
{(i)} tracking the use of AI systems,
and
{(ii)} keeping track of information related to model provenance (eg.~code and training data)~\cite{Gebru_2021,Mitchell_2019}.
Both aspects were also seen as useful for improving transparency and accountability,
which in turn can be useful for building trust.

Usage tracking was additionally seen as helpful for evaluating the effect of user interventions on system performance:
\textit{``[the system] suggested doing one scenario, we chose to do another,
this is the result we got [...] did we do the job that we expected?
Or did we do the job that the system expected?''} (X15).

Keeping logs and previous versions of data/models/systems was suggested:
\textit{``When the system gets complex, you have to keep more evidence along the way.
Version control, and the immutable log. You don't want people to tamper this [...] after things went wrong''} (X02).
Many interviewees used established software development management tools to explicitly keep previous revisions;
for example: \textit{``Any software we are developing is in Bitbucket, internal configuration management system''} (X17).

\newpage
\renewcommand{\baselinestretch}{1.00}\small\normalsize

\section{Limitations}
\label{sec:limitations}

The interviewees for this investigation were selected through solicitation emails and recommendations,
which may pose a threat to internal validity.
While selection bias is a possibility when the interviewees are not randomly selected,
the threat is partially alleviated as the interviewers had no contact with the interviewees beforehand.
Moreover, the interviewees had various backgrounds, roles, and genders.

A saturation of findings was reached after interviewing 21 participants.
To reduce the risk of missing information and interviewer subjectivity,
each interview included three interviewers with diverse research backgrounds.
The interviewers worked jointly to pose questions,
which can aid in increasing the range and depth of inquiry,
as well as reducing the likelihood of subjective bias on the stopping point of questions.

This investigation was conducted within one organisation, which may pose a threat to external validity;
the opinions provided by the interviewees may not be representative of the larger AI development community.
To reduce this threat, we ensured that the interviewees had various roles and degrees of expertise,
and worked on a variety of research areas and projects (for both internal and external customers).

~

\section{Concluding Remarks}
\label{sec:conclusion}

Existing AI ethics principles are typically high-level
and do not provide tangible advice on how to design and develop responsible AI systems.
In this work we have presented an empirical investigation
with the aim of increasing the understanding of practitioners' perceptions of AI ethics principles,
as well as their possible implementation.

A recent interview study involving high-level organisational leaders
provides empirical evidence that the following four sets of practices are likely to be required
for the implementation of AI ethics principles~\cite{Seppala_2021}:
(i)~governance,
(ii)~{AI}~system design and development,
(iii)~competence and knowledge development,
and
(iv)~stakeholder communication.

The investigation presented here specifically focuses on the problem of implementing high-level AI principles
through the lens of AI system design and development,
adapting processes used in software engineering.
The salient findings cover four aspects:
(i)~high-level view,
(ii)~requirements engineering,
(iii)~design and implementation,
(iv)~deployment and operation.

The observations and comments given by the interviewees for this investigation
provide insights into the challenges that practitioners are facing when dealing with AI ethics issues
during research, development and deployment. 
The suggestions presented in this work are necessarily not exhaustive,
as they reflect the content of the interviews and the surrounding discussions.
However, this work complements recent literature such as \cite{Mantymaki_2022,Morley_2020,Raji_2020b,Shneiderman_2020,Shneiderman_2021,Smit_2020},
which allows a more complete picture to be obtained on the translation of AI ethics principles into practice.

\textcolor{white}
{
\fontsize{1}{1}\selectfont
It must be also noted that system design and development
can be interpreted as one component within a larger framework for implementing AI ethics principles.
A recent interview study involving high-level organisational leaders
provides empirical evidence that the following four sets of practices may be required
for the implementation of such principles~\cite{Seppala_2021}:
(i)~governance,
(ii)~{AI} design and development,
(iii)~competence and knowledge development,
and
(iv)~stakeholder communication.
}

\balance
\bibliographystyle{ieee_mod}
\bibliography{references}

\begin{thebibliography}{10}\itemsep=1ex

\bibitem{DISER_2020}
{Australian Government}.
\newblock Australia's artificial intelligence ethics framework, 2019.
\newblock Online.
  https://industry.gov.au/data-and-publications/australias-artificial-intelligence-ethics-framework.

\bibitem{Arrieta_2020}
A.~Barredo~Arrieta, N.~Díaz-Rodríguez, J.~Del~Ser, A.~Bennetot, S.~Tabik,
  A.~Barbado, S.~Garcia, S.~Gil-Lopez, D.~Molina, R.~Benjamins, R.~Chatila, and
  F.~Herrera.
\newblock Explainable artificial intelligence ({XAI}): Concepts, taxonomies,
  opportunities and challenges toward responsible {AI}.
\newblock {\em Information Fusion}, 58:82--115, 2020.

\bibitem{jan2019}
J.~Bosch.
\newblock From efficiency to effectiveness: Delivering business value through
  software.
\newblock In S.~Hyrynsalmi, M.~Suoranta, A.~Nguyen-Duc, P.~Tyrväinen, and
  P.~Abrahamsson, editors, {\em Software Business}, pages 3--10. Springer
  International Publishing, 2019.

\bibitem{Braun_2006}
V.~Braun and V.~Clarke.
\newblock Using thematic analysis in psychology.
\newblock {\em Qualitative Research in Psychology}, 3(2):77--101, 2006.

\bibitem{Christoforaki_2022}
M.~Christoforaki and O.~Beyan.
\newblock {AI} ethics -- a bird's eye view.
\newblock {\em Applied Sciences}, 12(9):4130, 2022.

\bibitem{Colvin_2016}
R.~Colvin, G.~B. Witt, and J.~Lacey.
\newblock Approaches to identifying stakeholders in environmental management:
  Insights from practitioners to go beyond the `usual suspects`.
\newblock {\em Land Use Policy}, 52:266--276, 2016.

\bibitem{Eubanks_2019}
V.~Eubanks.
\newblock {\em Automating Inequality: How High-Tech Tools Profile, Police, and
  Punish the Poor}.
\newblock Picador, New York, NY, 2019.

\bibitem{fjeld2020principled}
J.~Fjeld, N.~Achten, H.~Hilligoss, A.~C. Nagy, and M.~Srikumar.
\newblock Principled artificial intelligence: Mapping consensus in ethical and
  rights-based approaches to principles for {AI}.
\newblock {Research Publication No.~2020-1}, Berkman Klein Center for Internet
  \& Society at Harvard University, 2020.

\bibitem{friedman_bias_1996}
B.~Friedman and H.~Nissenbaum.
\newblock Bias in {Computer} {Systems}.
\newblock {\em {ACM} Transactions on Computer Systems}, 14(3):330--347, 1996.

\bibitem{Gebru_2021}
T.~Gebru, J.~Morgenstern, B.~Vecchione, J.~W. Vaughan, H.~Wallach,
  H.~{Daum\'{e}~III}, and K.~Crawford.
\newblock Datasheets for datasets.
\newblock {\em Communications of the {ACM}}, 64(12):86--92, 2021.

\bibitem{Hajkowicz_2023}
S.~Hajkowicz, C.~Sanderson, S.~Karimi, A.~Bratanova, and C.~Naughtin.
\newblock The diffusion of artificial intelligence technology across research
  fields: A bibliometric analysis of scholarly publications from 1960-2021.
\newblock {\em SSRN}, 2023.
\newblock {DOI:}
  \href{https://doi.org/10.2139/ssrn.4330250}{\textcolor{black}{10.2139/ssrn.4330250}}.

\bibitem{Hall_2015}
N.~Hall, J.~Lacey, S.~Carr-Cornish, and A.-M. Dowd.
\newblock Social licence to operate: understanding how a concept has been
  translated into practice in energy industries.
\newblock {\em Journal of Cleaner Production}, 86:301--310, 2015.

\bibitem{jobin2019global}
A.~Jobin, M.~Ienca, and E.~Vayena.
\newblock The global landscape of {AI} ethics guidelines.
\newblock {\em Nature Machine Intelligence}, 1(9):389--399, 2019.

\bibitem{johnson_ethics_2009}
D.~G. Johnson.
\newblock {\em Computer Ethics}.
\newblock Pearson Education, 4th edition, 2009.

\bibitem{Mantymaki_2022}
M.~M\"{a}ntym\"{a}ki, M.~Minkkinen, T.~Birkstedt, and M.~Viljanen.
\newblock Defining organizational {AI} governance.
\newblock {\em AI and Ethics}, 2(4):603--609, 2022.

\bibitem{Mitchell_2019}
M.~Mitchell, S.~Wu, A.~Zaldivar, P.~Barnes, L.~Vasserman, B.~Hutchinson,
  E.~Spitzer, I.~D. Raji, and T.~Gebru.
\newblock Model cards for model reporting.
\newblock In {\em Conference on Fairness, Accountability, and Transparency},
  pages 220--229. ACM, 2019.

\bibitem{Mittelstadt_2019}
B.~Mittelstadt.
\newblock Principles alone cannot guarantee ethical {AI}.
\newblock {\em Nature Machine Intelligence}, 1(11):501--507, Nov 2019.

\bibitem{Morley_2020}
J.~Morley, L.~Floridi, L.~Kinsey, and A.~Elhalal.
\newblock From what to how: An initial review of publicly available {AI} ethics
  tools, methods and research to translate principles into practices.
\newblock {\em Science and Engineering Ethics}, 26(4):2141--2168, 2020.

\bibitem{NMHRC_2018}
{National Health and Medical Research Council}, {Australian Research Council},
  and {Universities Australia}.
\newblock {National Statement on Ethical Conduct in Human Research (Updated)},
  2018.
\newblock Online.
  https://www.nhmrc.gov.au/about-us/publications/national-statement-ethical-conduct-human-research-2007-updated-2018.

\bibitem{Paul_2008}
C.~L. Paul.
\newblock A modified delphi approach to a new card sorting methodology.
\newblock {\em Journal of Usability Studies}, 4(1):7--30, 2008.

\bibitem{Raji_2020b}
I.~D. Raji, A.~Smart, R.~N. White, M.~Mitchell, T.~Gebru, B.~Hutchinson,
  J.~Smith-Loud, D.~Theron, and P.~Barnes.
\newblock Closing the {AI} accountability gap: Defining an end-to-end framework
  for internal algorithmic auditing.
\newblock In {\em Conference on Fairness, Accountability, and Transparency},
  2020.

\bibitem{Sanderson_2023}
C.~Sanderson, D.~Douglas, Q.~Lu, E.~Schleiger, J.~Whittle, J.~Lacey,
  G.~Newnham, S.~Hajkowicz, C.~Robinson, and D.~Hansen.
\newblock \scalebox{0.9665}{{AI} ethics principles in practice: perspectives of
  designers and developers}.
\newblock {\em IEEE Transactions on Technology and Society}, 2023.
\newblock
  {DOI:}~\href{https://doi.org/10.1109/TTS.2023.3257303}{\textcolor{black}{10.1109/TTS.2023.3257303}}.

\bibitem{google_technical_debt}
D.~Sculley, G.~Holt, D.~Golovin, E.~Davydov, T.~Phillips, D.~Ebner,
  V.~Chaudhary, M.~Young, J.-F. Crespo, and D.~Dennison.
\newblock Hidden technical debt in machine learning systems.
\newblock In {\em Advances in Neural Information Processing Systems},
  volume~28, pages 2503--2511, 2015.

\bibitem{Seppala_2021}
A.~Sepp\"{a}l\"{a}, T.~Birkstedt, and M.~M\"{a}ntym\"{a}ki.
\newblock From ethical {AI} principles to governed {AI}.
\newblock In {\em International Conference on Information Systems}, 2021.

\bibitem{Shneiderman_2020}
B.~Shneiderman.
\newblock Bridging the gap between ethics and practice: Guidelines for
  reliable, safe, and trustworthy human-centered {AI} systems.
\newblock {\em ACM Transactions on Interactive Intelligent Systems},
  10(4):1--31, 2020.

\bibitem{Shneiderman_2021}
B.~Shneiderman.
\newblock Responsible {AI}: Bridging from ethics to practice.
\newblock {\em Communications of the ACM}, 64(8):32--35, 2021.

\bibitem{Smit_2020}
K.~Smit, M.~Zoet, and J.~van Meerten.
\newblock A review of {AI} principles in practice.
\newblock In {\em Proc. Pacific Asia Conference on Information Systems}, 2020.

\bibitem{Tigard2021}
D.~W. Tigard.
\newblock Responsible {AI} and moral responsibility: a common appreciation.
\newblock {\em {AI} and Ethics}, 1(2):113--117, 2021.

\bibitem{Whittle19}
J.~Whittle.
\newblock Is your software valueless?
\newblock {\em {IEEE} Software}, 36(3):112--115, 2019.

\bibitem{Whittlestone_2019}
J.~Whittlestone, R.~Nyrup, A.~Alexandrova, and S.~Cave.
\newblock The role and limits of principles in {AI} ethics: Towards a focus on
  tensions.
\newblock In {\em AAAI/ACM Conference on AI, Ethics, and Society}, 2019.

\bibitem{zhang21}
B.~Zhang, M.~Anderljung, L.~Khan, N.~Dreksler, M.~C. Horowitz, and A.~Dafoe.
\newblock Ethics and governance of artificial intelligence: Evidence from a
  survey of machine learning researchers.
\newblock {\em Journal of Artificial Intelligence Research}, 71:591--666, 2021.

\end{thebibliography}

\end{document}